# LEPTON TEXTURES AND NEUTRINO OSCILLATIONS

ROHIT VERMA

*Rayat Institute of Engineering and Information Technology,*
*Near Ropar, India*
rohitverma@live.com

Systematic analyses of the textures arising in lepton mass matrices have been carried out using unitary transformations and condition of naturalness for the Dirac and Majorana neutrino possibilities. It is observed that the recent three neutrino oscillation data together with the effective mass in neutrinoless double beta decay provide vital clues in predicting the general structures of these lepton mass matrices.

## 1. Introduction

The recent measurements of large[1] $\theta_{13}$ along with indications of a non-maximal[2] $\theta_{23}$ point that the masses and flavor mixing patterns for quarks and leptons are significantly different. Whereas for the case of quarks, one observes large mass hierarchy, small mixing angles with relatively large masses, however, for the case of leptons, neutrino mass spectrum is yet undetermined, there are two large mixing angles and these involve relatively small neutrino masses. Neutrino physics has yet to address several intriguing phenomenon like the origin of small neutrino masses, the absolute neutrino mass scale, the neutrino mass hierarchy i.e. normal or inverted, the nature of neutrinos i.e. Dirac or Majorana and the possibility of lepton CP violation. This makes the task of constructing the corresponding mass matrices for quarks and leptons more challenging, especially if the fermion mass matrices are to be considered in a unified framework[3].

Due to the large redundancy in fermion mass matrices in SM framework, these matrices are arbitrary complex matrices involving 36 free parameters, way large as compared to the number of physical observables. This redundancy is related to the fact that one has the freedom to make unitary transformations under which the fermion mass matrices change but the gauge currents remain real and diagonal.

The problem of constructing the lepton mass matrices may be addressed using the Bottom Up approach which involves using the experimental data to extract clues for the structure of fermion mass matrices compatible with the observed fermion masses and mixing patterns. One such approach is the 'texture zero' approach initiated by Weinberg[4] and Fritzsch[5]. *A particular texture structure is said to be texture n zero, if the sum of the number of diagonal zeros and half the number of the symmetrically placed off diagonal zeros is n.* Interestingly in this approach, it becomes important to investigate when a set of texture zeros result only from the choice of a given basis and when these imply restrictions on fermion mass matrices. Recently[6,7], it has been shown that some sets of texture zeros are physically insignificant since these can be obtained starting from arbitrary fermion mass matrices by making appropriate unitary transformations also called Weak Basis (WB) transformations. Such transformations allow one to obtain

Hermitian fermion mass matrices involving a 'maximum' of three phenomenological texture zeros. Any 'additional' texture zero is supposed to have physical implications.

In the current paper, we systematically study the lepton textures arising from WB transformations, both for the case of Dirac as well as Majorana neutrinos and attempt to derive invariant relations for lepton mixing angles assuming normal neutrino mass hierarchy compatible with natural structures of lepton mass matrices. We also attempt to gain an insight on the maximum number of texture zeros that may be incorporated in these mass matrices without adding any physical implications. The significance of the effective mass in neutrinoless double beta decay 0νββ decay in determining the texture structure of these mass matrices is also addressed.

## 2. Naturalness and Unitary Transformations

In the WB approach, one may consider a basis wherein for example,

$$M_e = D_e, \quad M_{\nu D} = V D_{\nu D} V^\dagger = \begin{pmatrix} e_{\nu D} & |a_{\nu D}|e^{i\alpha_{\nu D}} & |f_{\nu D}|e^{i\omega_{\nu D}} \\ |a_{\nu D}|e^{-i\alpha_{\nu D}} & d_{\nu D} & |b_{\nu D}|e^{i\beta_{\nu D}} \\ |f_{\nu D}|e^{-i\omega_{\nu D}} & |b_{\nu D}|e^{-i\beta_{\nu D}} & c_{\nu D} \end{pmatrix}. \quad (1)$$

Here the charged lepton mass matrix $M_e$ is real diagonal such that $D_e$ = diag ($m_e$, -$m_\mu$, $m_\tau$) while the Dirac neutrino mass matrix $M_{\nu D}$ is an arbitrary Hermitian mass matrix with $D_{\nu D}$ = diag ($m_{\nu 1D}$, -$m_{\nu 2D}$, $m_{\nu 3D}$) and V is the neutrino mixing matrix also called Pontecorvo-Maki-Nakagawa-Sakata (PMNS) matrix[8]. For the quark sector, we observe that the hierarchy among the quark masses i.e. $m_1 \ll m_2 \ll m_3$ and the mixing matrix U elements i.e. $V_{13,31} < V_{23,32} < V_{12,21} < V_{11,22,33}$ is naturally translated on the corresponding mass matrices i.e.

$$e < (|a|, |f|) < d < |b| < c. \quad (2)$$

Such hierarchical mass matrices have been referred to in the literature as *natural mass matrices*[9]. In principle, an exact diagonalization of the mass matrix given in Eq. (1) is not always possible. In this context, one can apply a WB transformation[6,7] U on the mass matrices $M_e$ and $M_{\nu D}$, such that

$$M_e \to M'_e = U M_e U^\dagger, \quad M_{\nu D} \to M'_{\nu D} = U M_{\nu D} U^\dagger. \quad (3)$$

It is trivial to check that the two representations $(M_e, M_{\nu D})$ and $(M'_e, M'_{\nu D})$ are physically equivalent leading to the same mixing matrix, provided neutrinos are assumed to be Dirac particles. In such a case[7], there exists a possible choice of U such that

$$(M'_e)_{13,31} = (M'_{\nu D})_{13,31} = (M'_{\nu D})_{11} = 0, \quad (4)$$

with non-vanishing other elements. However, in case the neutrinos are of Majorana type, the light neutrino mass matrix is obtained using the seesaw mechanism[10] as $M'_\nu = -M'^T_{\nu D} M_R^{-1} M'_{\nu D}$, where $M_R$ is the right handed Majorana mass matrix. It can be seen[11] that for Hermitian $M'_e$ and $M'_{\nu D}$ and for real diagonal $M_R = m_R I$, where I is a unit matrix and $m_R$ denotes a very large mass scale, the Left diagonalizing transformations for $M'_{\nu D}$ and $M'_\nu$ remain the same so that the light neutrino mass matrix is given by

$$M'_\nu = -M'^T_{\nu D} M_R^{-1} M'_{\nu D} = P_{\nu D} O_{\nu D} \frac{(D_{\nu D})^2}{m_R} O^T_{\nu D} P_{\nu D} = P_{\nu D} O_{\nu D} D_\nu O^T_{\nu D} P_{\nu D}. \quad (5)$$

where $O_{\nu D}$ is the diagonalizing transformation for matrix $M'_{\nu D}$, defined through
$$D_{\nu D} = O^{\dagger}_{\nu D} P_{\nu D} M'_{\nu D} Q_{\nu D} O_{\nu D} \qquad (6)$$
where $P_{\nu D} = \mathrm{diag}(e^{-i\alpha_{\nu D}}, 1, e^{i\beta_{\nu D}})$ with $Q_e = P^{\dagger}_e$ (for Hermitian $M'_{\nu D}$). Such a simple $M_R$ structure allows the two representations $(M'_e, M'_{\nu D})$ and $(M'_e, M'_{\nu D}, M'_{\nu})$ to be physically equivalent.

## 3. Matrix Diagonalization

For the WB textures of lepton mass matrices obtained through "Eq. (3) and (4)", $M_{\nu D}$ emerges as a Fritzsch-like texture two zero matrix (with $e = 0$) whereas $M_e$ has the following structure

$$M'_e = \begin{pmatrix} e_e & |a_e|e^{i\alpha_e} & 0 \\ |a_e|e^{-i\alpha_e} & d_e & |b_e|e^{i\beta_e} \\ 0 & |b_e|e^{-i\beta_e} & c_e \end{pmatrix}. \qquad (7)$$

The exact diagonalization of the matrix $M'_e$ can then be carried out using the three matrix invariants viz. Trace $M'_e$, Trace $(M'_e)^2$ and Determinant $M'_e$. These provide the following equations relating the mass matrix elements $c_e$, $a_e$ and $b_e$ with the charged lepton masses $m_1 = m_e$, $m_2 = m_\mu$, $m_3 = m_\tau$ and the free parameters $d_e$ and $e_e$, e.g.,

$$c_e = m_1 - m_2 + m_3 - d_e - e_e, \qquad (8a)$$

$$|a_e| = \sqrt{\frac{(m_1 - e_e)(m_2 + e_e)(m_3 - e_e)}{(c_e - e_e)}}, \qquad (8b)$$

$$|b_e| = \sqrt{\frac{(c_e - m_1)(m_3 - c_e)(c_e + m_2)}{(c_e - e_e)}}. \qquad (8c)$$

It is observed[12] that for $|a_e|$ and $|b_e|$ to remain real, the free parameters $d_e$ and $e_e$ get constrained within the limits

$$m_1 > e_e > -m_2, \quad (m_3 - m_2 - e_e) > d_e > (m_1 - m_2 - e_e). \qquad (9)$$

The above constraints indicate that the condition of Hermicity on the texture one zero mass matrix in "Eq. (7)" restrict the free parameter $e_e$ to have small values only, consistent with the naturalness condition in "Eq. (2)". The exact diagonalizing matrix $O_e$ for $M'_e$ defined through

$$D_e = O^{\dagger}_e P_e M'_e Q_e O_e \qquad (10)$$

where $O_e =$

$$\begin{pmatrix} \sqrt{\dfrac{(e_e+m_2)(m_3-e_e)(c_e-m_1)}{(c_e-e_e)(m_3-m_1)(m_2+m_1)}} & \sqrt{\dfrac{(m_1-e_e)(m_3-e_e)(c_e+m_2)}{(c_e-e_e)(m_3+m_2)(m_2+m_1)}} & \sqrt{\dfrac{(m_1-e_e)(e_e+m_2)(m_3-c_e)}{(c_e-e_e)(m_3+m_2)(m_3-m_1)}} \\ \sqrt{\dfrac{(m_1-e_e)(c_e-m_1)}{(m_3-m_1)(m_2+m_1)}} & -\sqrt{\dfrac{(e_e+m_2)(c_e+m_2)}{(m_3+m_2)(m_2+m_1)}} & \sqrt{\dfrac{(m_3-e_e)(m_3-c_e)}{(m_3+m_2)(m_3-m_1)}} \\ -\sqrt{\dfrac{(m_1-e_e)(m_3-c_e)(c_e+m_2)}{(c_e-e_e)(m_3-m_1)(m_2+m_1)}} & \sqrt{\dfrac{(e_e+m_2)(c_e-m_1)(m_3-c_e)}{(c_e-e_e)(m_3+m_2)(m_2+m_1)}} & \sqrt{\dfrac{(m_3-e_e)(c_e-m_1)(c_e+m_2)}{(c_e-e_e)(m_3+m_2)(m_3-m_1)}} \end{pmatrix}. (11)$$

Likewise, the diagonalizing transformation $O_{vD}$ for the Dirac neutrino matrix $M'_{vD}$ can be obtained by substituting $e = 0$ in the above transformations, so that

$$O_{vD} = \begin{pmatrix} \sqrt{\dfrac{m_2 m_3 (c_{vD} - m_1)}{c_{vD}(m_3 - m_1)(m_2 + m_1)}} & \sqrt{\dfrac{m_1 m_3 (c_{vD} + m_2)}{c_{vD}(m_3 + m_2)(m_2 + m_1)}} & \sqrt{\dfrac{m_1 m_2 (m_3 - c_{vD})}{c_{vD}(m_3 + m_2)(m_3 - m_1)}} \\ \sqrt{\dfrac{m_1(c_{vD} - m_1)}{(m_3 - m_1)(m_2 + m_1)}} & -\sqrt{\dfrac{(e_e + m_2)(c_e + m_2)}{(m_3 + m_2)(m_2 + m_1)}} & \sqrt{\dfrac{(m_3 - e_e)(m_3 - c_e)}{(m_3 + m_2)(m_3 - m_1)}} \\ -\sqrt{\dfrac{m_1(m_3 - c_{vD})(c_{vD} + m_2)}{c_{vD}(m_3 - m_1)(m_2 + m_1)}} & \sqrt{\dfrac{m_2(c_{vD} - m_1)(m_3 - c_{vD})}{c_{vD}(m_3 + m_2)(m_2 + m_1)}} & \sqrt{\dfrac{m_3(c_{vD} - m_1)(c_{vD} + m_2)}{c_{vD}(m_3 + m_2)(m_3 - m_1)}} \end{pmatrix}. \quad (12)$$

One can now easily compute the PMNS matrix for the cases of Dirac as well as Majorana neutrinos through

$$V = O_e^\dagger Q_e P_{vD} O_{vD}. \quad (13)$$

in agreement with "Eqns. (5), (6) and (10)". In general,

$$V_{i\sigma} = O_{1i}^e O_{1\sigma}^v e^{-i\phi_1} + O_{2i}^e O_{2\sigma}^v + O_{3i}^e O_{3\sigma}^v e^{i\phi_2}, \quad (14)$$

where the phases $\phi_1 = \alpha_e - \alpha_{vD}$ and $\phi_2 = \beta_e - \beta_{vD}$ are also free parameters.

However, the expressions for the neutrino mixing angles computed using the "Eqns. (11) - (14)" are quite lengthy and difficult to comprehend especially if one intends to study the implications of the various elements of the mass matrices for the neutrino mixing angles. To this end, one may adopt a more convenient parameterization by redefining the diagonal free parameters in the charged lepton mass matrix $M'_e$ as $\xi_e$ and $\zeta_e$ defined as $\xi_e = e_e/m_e$ and $\zeta_e = d_e/c_e$ while $m_{e\mu} = m_e/m_\mu$, $m_{e\tau} = m_e/m_\tau$ along with $m_{\mu\tau} = m_\mu/m_\tau$ have may be considered for simplicity. Likewise for the Dirac neutrino matrix, one may define these hierarchy characterizing parameters as $\xi_{vD} = e_{vD}/m_{v1D}$, $\zeta_{vD} = d_{vD}/c_{vD}$ while $m_{v12D} = m_{v1D}/m_{v2D}$, $m_{v13D} = m_{v1D}/m_{v3D}$ and $m_{v23D} = m_{v2D}/m_{v3D}$ have again been considered for simplicity. This greatly simplifies the diagonalizing transformation matrices for the lepton mass matrices, which assume the following form[13]

$$O_e = \begin{pmatrix} 1 & \sqrt{\dfrac{m_{e\mu}(1-\xi_e)}{(1+m_{\mu\tau})}} & \sqrt{\dfrac{m_{e\tau}m_{\mu\tau}(\zeta_e + m_{\mu\tau})(1-\xi_e)}{(1+m_{\mu\tau})}} \\ \sqrt{\dfrac{m_{e\mu}(1-\xi_e)}{(1+\zeta_e)}} & -\sqrt{\dfrac{1}{(1+\zeta_e)(1+m_{\mu\tau})}} & \sqrt{\dfrac{(\zeta_e + m_{\mu\tau})}{(1+\zeta_e)(1+m_{\mu\tau})}} \\ -\sqrt{\dfrac{m_{e\mu}(\zeta_e + m_{\mu\tau})(1-\xi_e)}{(1+\zeta_e)}} & \sqrt{\dfrac{(\zeta_e + m_{\mu\tau})}{(1+\zeta_e)(1+m_{\mu\tau})}} & \sqrt{\dfrac{1}{(1+\zeta_e)(1+m_{\mu\tau})}} \end{pmatrix}, \quad (15)$$

$$O_{\nu D} = \begin{pmatrix} \sqrt{\dfrac{(1+\xi_{\nu D} m_{\nu 12D})}{(1+m_{\nu 12D})}} & \sqrt{\dfrac{m_{\nu 12D}(1-\xi_{\nu D})}{(1+m_{\nu 12D})(1+m_{\nu 23D})}} & \kappa\sqrt{\dfrac{m_{\nu 13D} m_{\nu 23D}(m_{\nu 23D}+\zeta_{\nu D})(1+\xi_{\nu D} m_{\nu 12D})}{(1+m_{\nu 23D})(1+m_{\nu 23D})}} \\ \sqrt{\dfrac{m_{\nu 12D}(1-\xi_{\nu D})(1+m_{\nu 23D})}{(1+m_{\nu 12D})(1+\zeta_{\nu D})}} & -\sqrt{\dfrac{(1+\xi_{\nu D} m_{\nu 12D})}{(1+m_{\nu 12D})(1+\zeta_{\nu D})}} & \sqrt{\dfrac{(m_{\nu 23D}+\zeta_{\nu D})}{(1+m_{\nu 23D})(1+\zeta_{\nu D})}} \\ -\sqrt{\dfrac{m_{\nu 12D}(m_{\nu 23D}+\zeta_{\nu D})}{(1+m_{\nu 12D})(1+\zeta_{\nu D})}} & \sqrt{\dfrac{(1+\xi_{\nu D} m_{\nu 12D})(m_{\nu 23D}+\zeta_{\nu D})}{(1+m_{\nu 12D})(1+m_{\nu 23D})(1+\zeta_{\nu D})}} & \sqrt{\dfrac{1}{(1+\zeta_{\nu D})}} \end{pmatrix}, \quad (16)$$

where $\kappa = \sqrt{(1-\xi_{\nu D})/(1-m_{\nu 13})}$. The condition of naturalness has been imposed on the lepton mass matrices $(M'_e, M'_{\nu D})$ by restricting the parameter spaces of the free parameters to $(\zeta_e, \zeta_{\nu D}, \xi_e, \xi_{\nu D}) < 1$ and assuming NH for the Dirac neutrinos i.e. $m_{\nu 1D} < m_{\nu 2D} < m_{\nu 3D}$.

## 4. PMNS Matrix

In case neutrinos are considered to be Dirac particles, they can acquire masses exactly in the same way as quarks and charged leptons do in the standard model. In this context, it has been shown that the highly-suppressed Yukawa couplings for Dirac neutrinos can naturally be achieved in the models with extra spacial dimensions[14] or through radiative mechanisms[15]. Noting that the possibility of Dirac neutrinos may still be allowed by the experiments[16] and that the inverted neutrino mass hierarchy appears to be ruled out for several lepton mass matrices[17], the PMNS matrix for the case of Dirac neutrinos may be obtained through "Eq. 13" and the resulting simplified expressions for the three lepton mixing angles are obtained as

$$s_{12} = \sqrt{\dfrac{m_{\nu 12D}}{(1+m_{\nu 12D})(1+m_{\nu 23D})}}, \quad (17)$$

$$s_{13} = \left| \sqrt{\dfrac{m_{\nu 13D} m_{\nu 23D}(m_{\nu 23D}+\zeta_{\nu D})}{(1+m_{\nu 23D})(1-m_{\nu 13D})(1+m_{\nu 23D})}} e^{-i\phi_1} - \sqrt{\dfrac{m_{e\mu}(1-\xi_e)}{(1+\zeta_e)(1+m_{\nu 23D})(1+\zeta_{\nu D})}}\left(\sqrt{(m_{\nu 23D}+\zeta_{\nu D})} - \sqrt{(\zeta_e + m_{\mu\tau})(1+m_{\nu 23D})} e^{i\phi_2}\right) \right|, \quad (18)$$

$$s_{23} = \left| \sqrt{\dfrac{1}{(1+\zeta_e)(1+m_{\mu\tau})(1+m_{\nu 23D})(1+\zeta_{\nu D})}}\left(\sqrt{(m_{\nu 23D}+\zeta_{\nu D})} - \sqrt{(\zeta_e + m_{\mu\tau})(1+m_{\nu 23D})} e^{i\phi_2}\right) \right|. \quad (19)$$

Here only the leading order term (first) and the next to leading order terms have been retained. It is observed that the above relations hold good within an error of less than a percent. Note that the mixing angle $s_{12}$ depends only on the neutrino mass ratios $m_{\nu 12D}$ and $m_{\nu 23D}$. Likewise, it is also observed that the mixing angle $s_{23}$ is independent of $\xi_e$. This is easy to interpret as $\xi_e$ does not invoke any mixing among the second and the third generations of leptons. As a result, it should be interesting to investigate the implications of $\xi_e$, if any, for $s_{13}$ as well as those of $\zeta_e$ and $\zeta_{\nu D}$ for $s_{13}$ and $s_{23}$.

However, for the case of Majorana neutrinos, in order that the mixing angles are independent of $m_R$, we consider the neutrino masses $m_{v1}$, $m_{v2}$ and $m_{v3}$ as input parameters, so that we make the replacement of $m_{vD}$ with $\sqrt{m_v m_R}$ in all the terms of $O_{vD}$ in agreement with "Eq. (5)" above. In such a case the ratios $m_{v12D}$, $m_{v13D}$ and $m_{v23D}$ get replaced by $m_{v12D} \to \sqrt{m_{v12}} = \sqrt{m_{v1}/m_{v2}}$ and so on. Likewise, the free parameters get redefined as $\xi_{vD} \to \xi_v = e_{vD}/\sqrt{m_{v1} m_R} = e_v/\sqrt{m_{v1}}$ and $\zeta_{vD} \to \zeta_v = \zeta_{vD}$, where $\xi_v < 1$ and $\zeta_v < 1$, are again arbitrary and in accordance with the condition of naturalness. For the lepton mass matrices $(M'_e, M'_{vD}, M'_v)$ given by the "Eqs. (4, 5, 6, 13)", the product $Q_e P_{vD}$ in "Eq. (13)", is a diagonal phase matrix given by $Q_e P_{vD} = \text{diag}(e^{-i\phi_1}, 1, e^{i\phi_2})$, where the phases $\phi_1 = \alpha_e - \alpha_{vD}$ and $\phi_2 = \beta_e - \beta_{vD}$ are also free parameters. The expressions for the three lepton mixing angles may be expresses as

$$s_{12} = \sqrt{\frac{\sqrt{m_{v12}}}{\left(1+\sqrt{m_{v12}}\right)\left(1+\sqrt{m_{v23}}\right)}}, \qquad (20)$$

$$s_{13} = \left| \sqrt{\frac{\sqrt{m_{v13}}\sqrt{m_{v23}}\left(\sqrt{m_{v23}}+\zeta_v\right)}{\left(1+\sqrt{m_{v23}}\right)\left(1-\sqrt{m_{v13}}\right)\left(1+\sqrt{m_{v23}}\right)}} e^{-i\phi_1} - \sqrt{\frac{m_{e\mu}(1-\xi_e)}{(1+\zeta_e)(1+\sqrt{m_{v23}})(1+\zeta_v)}} \left( \sqrt{\left(\sqrt{m_{v23}}+\zeta_v\right)} - \sqrt{\left(\zeta_e+m_{\mu\tau}\right)\left(\sqrt{m_{v23}}+1\right)} e^{i\phi_2} \right) \right|, \qquad (21)$$

$$s_{23} = \left| \sqrt{\frac{1}{(1+\zeta_e)(1+m_{\mu\tau})(1+\sqrt{m_{v23}})(1+\zeta_v)}} \left( \sqrt{\left(\sqrt{m_{v23}}+\zeta_v\right)} - \sqrt{\left(\zeta_e+m_{\mu\tau}\right)\left(\sqrt{m_{v23}}+1\right)} e^{i\phi_2} \right) \right|. \qquad (22)$$

Again, one observes that the mixing angle $(s_{12})^2$ depends predominantly on the neutrino mass ratio $\sqrt{m_{v12}}$ and the mixing angle $s_{23}$ is still independent of the parameter $\xi_e$. As a result, it should be interesting to investigate the implications of $\xi_e$, if any, for $s_{13}$ as well as those of $\zeta_e$ and $\zeta_v$ for $s_{13}$ and $s_{23}$.

## 5. Inputs

The following 1σ C.L. values for the various three neutrino mixing parameters[18] have been used as inputs for the analysis, i.e.

$$\delta m^2 = (7.32 - 7.80) \times 10^{-5} \text{ GeV}^2,$$
$$\Delta m^2 = (2.33 - 2.49) \times 10^{-3} \text{ GeV}^2,$$
$$\sin^2 \theta_{12} = 0.29 - 0.33,$$
$$\sin^2 \theta_{13} = 0.022 - 0.027,$$
$$\sin^2 \theta_{23} = 0.37 - 0.41. \qquad (23)$$

Here the neutrino mass square differences are defined as $\delta m^2 = m_{v2}^2 - m_{v1}^2$ and $\Delta m^2 = m_{v3}^2 - (m_{v1}^2 + m_{v2}^2)/2$ for NH[18]. The "Eqns. (17), (18), (20) and (21)" imply a clear constraint on the neutrino mass ratios $m_{v12}$ and $m_{v13}$ through the neutrino oscillation parameters $s_{12}$ and $s_{13}$, since

$$m_{v12} = \sqrt{\frac{m_{v1}^2}{m_{v1}^2 + \delta m^2}} \quad \text{and} \quad m_{v13} = \sqrt{\frac{m_{v1}^2}{m_{v1}^2 + \Delta m^2 + (\delta m^2)/2}} \,. \tag{24}$$

In addition, we have imposed the condition of naturalness on the lepton mass matrices through the constraints $(\zeta_e, \zeta_{vD}, \zeta_v, \xi_e, \xi_v, \xi_{vD}) < 1$ and assumed NH for the neutrino masses, consistent with the condition of naturalness. Furthermore, in the absence of any clues for CP violation in the lepton sector, the phases $\phi_1$ and $\phi_2$ have been given full variation from 0 to $2\pi$.

## 6. Results

For the Dirac neutrino case, it is observed that the complete $1\sigma$ range of all the neutrino oscillation parameters given in "Eq. (23)" can be reconstructed by the relations (17-19). Furthermore, the complete range of the free parameters $\xi_e$ and $\zeta_e$, allowed by the condition of naturalness, does not seem to play a significant role in predicting the three mixing angles as depicted in Fig. 1.

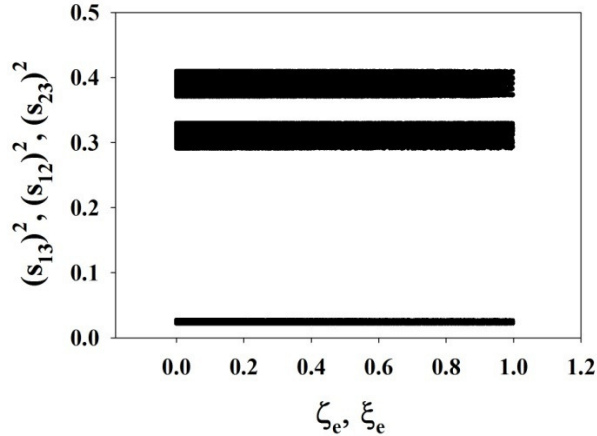

Fig. 1: Plot showing no dependence of mixing angles on $\xi_e$ and $\zeta_e$ in case of Dirac neutrinos.

One finds that the free parameters $\xi_e$ and $\zeta_e$ are completely redundant in predicting the three lepton mixing angles and hence may also be considered to be zero, in generality. As a result one may impose additional texture zeros at the (1,1) and (2,2) positions in the charged lepton mass matrix $M_e'$, over and above those imposed by the WB transformations of "Eq. (4)", without adding any physical implications for lepton mixing. However, the same is not observed for the free parameter $\zeta_{vD}$, as shown in Fig. 2. It is observed that the values of $\zeta_{vD} < 0.36$ are not able to reproduce the mixing angles $s_{13}$ and $s_{23}$. For these small values of $\zeta_{vD}$, the leading order terms in $s_{13}$ and $s_{23}$ are not able to regenerate the corresponding experimentally allowed values, implying that $\zeta_{vD} = 0$ does have physical implications for neutrino oscillation phenomenology. As a result, for the case of Dirac neutrinos, the most general texture three zero lepton mass matrices of "Eq. (4)", obtained through WB transformations, are physically equivalent to texture five zero Hermitian lepton mass matrices with $\xi_e = 0$, $\xi_v = 0$, $\zeta_e = 0$ and $\zeta_v \neq 0$, when the condition of naturalness is imposed on these. The corresponding mod values of the neutrino mixing matrix elements for such texture five zero Dirac lepton mass matrices are given by

$$V = \begin{bmatrix} 0.8076 - 0.8336 & 0.5314 - 0.5683 & 0.1449 - 0.1643 \\ 0.3309 - 0.4003 & 0.6711 - 0.7185 & 0.6001 - 0.6335 \\ 0.4108 - 0.4678 & 0.4275 - 0.4935 & 0.7757 - 0.7853 \end{bmatrix}, \qquad (25)$$

which are in good agreement with the current three neutrino oscillation data given by "Eq. (23)".

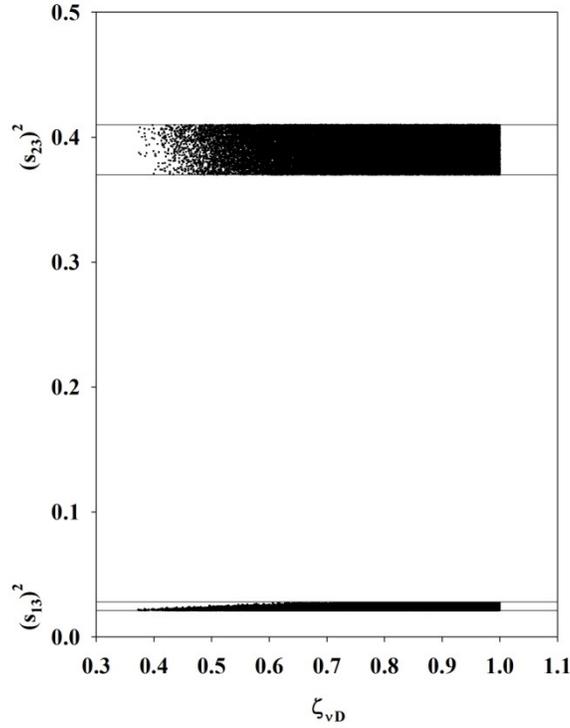

Fig. 2: Plot depicting the non redundancy of the free parameter $\zeta_{vD}$ for the Dirac Neutrino case. The horizontal lines represent the experimentally allowed values of the corresponding mixing angles.

Likewise, for the case of Majorana neutrinos, one observes that the relations (20-22) are able to reconstruct the complete $1\sigma$ range of all the neutrino oscillation parameters given in "Eq. (23)". Interestingly, all the free parameters $\xi_e$, $\zeta_v$ and $\zeta_e$, allowed by the condition of naturalness, appear to be redundant in this case. Whereas the plot of $\xi_e$ or $\zeta_e$ versus $s_{13}$, $s_{12}$ and $s_{23}$ is similar to Fig. 1, that of $\zeta_v$ versus $s_{13}$ and $s_{23}$ is depicted in Fig. 3 below. The presence of $\left(\sqrt{m_{v23}} + \zeta_v\right)$ in the leading order terms for $s_{23}$ and $s_{13}$, restricts the phenomenological range of $\zeta_v$ to $0 < \zeta_v < 0.70$ in order to regenerate the experimentally measured $s_{13}$. It is also observed that values of $\zeta_\nu > 0.70$ lead to an overshoot in $s_{13}$ from its experimental range.

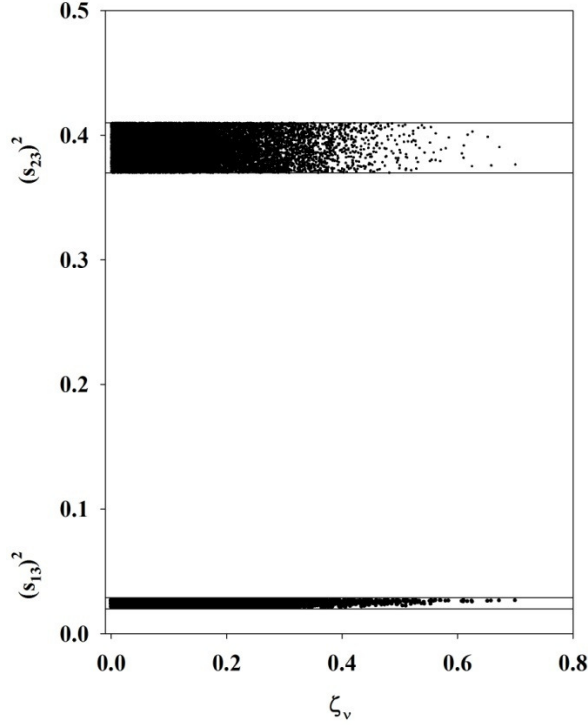

Fig. 3: Plot depicting the redundancy of the free parameter $\zeta_\nu$ for the Majorana Neutrino case. The horizontal lines represent the experimentally allowed values of the corresponding mixing angles.

From the point of view of three neutrino oscillation data, unlike the Dirac case, the lepton mass matrices for Majorana neutrinos may be reduced to texture six zero mass matrices with $\xi_e = 0$, $\xi_\nu = 0$, $\zeta_e = 0$ and $\zeta_\nu = 0$, without any loss of generality. Since the unique window to verify the Majorana nature of massive neutrinos is through the neutrinoless double beta (0νββ) decay, it is also desirable to study the impact of the parameters $\zeta_\nu$ on the effective mass $m_{ee}$ measured in (0νββ) defined through[19]

$$m_{ee} = m_{\nu 1} |V_{e1}|^2 + m_{\nu 2} |V_{e2}|^2 + m_{\nu 3} |V_{e3}|^2. \qquad (26)$$

Interestingly, the large values of the parameter $\zeta_\nu$ appear to have greater implications[20] on the allowed range of $m_{ee}$ and hence only the parameters $\xi_e$ and $\zeta_e$ may be considered to be completely redundant if neutrinos are Majorana particles. This further implies that even for the Majorana neutrino case, the maximum number of texture zeros that may be imposed on the corresponding lepton mass matrices are five characterized by $\zeta_e = 0$,

$\xi_e = 0$, $\xi_v = 0$, $\zeta_v \neq 0$. The corresponding mod values of the neutrino mixing matrix elements for such texture five zero lepton mass matrices are given by

$$V = \begin{bmatrix} 0.8075 - 0.8336 & 0.5313 - 0.5683 & 0.1449 - 0.1643 \\ 0.3291 - 0.4058 & 0.6707 - 0.7208 & 0.6001 - 0.6335 \\ 0.4032 - 0.4696 & 0.4265 - 0.4971 & 0.7578 - 0.7852 \end{bmatrix}, \qquad (27)$$

which are also in good agreement with the recent three neutrino oscillation data,

**Conclusions**

It is observed that, although the neutrino mixing pattern is significantly different from the quarks, yet it can also be described by 'natural' mass matrices. For Dirac neutrino case, it appears that the phenomenological difference between the three zero (WB choice) and five zero (two assumptions added) textures is physically insignificant under the condition of naturalness. For Majorana case, no new restrictions are imposed on neutrino mixing in going from three texture zeros to six texture zeros, if considerations of $m_{ee}$ are not taken into account. Otherwise, the lepton mass matrices reduce to texture five zeros, both for the Dirac as well as the Majorana cases, without any loss of generality. However, the symmetry that leads to such texture five zero structures requires a careful investigation in a unified framework.

**Acknowledgements**

I would like to thank the organizers of 'International Conference on Flavor Physics and Mass Generation', NTU Singapore, for providing an opportunity to present this work. Thanks are also due to the Director, Rayat Institute of Engineering and Information Technology, for providing the necessary working facilities.